\documentclass[pra,aps]{revtex4}
\usepackage{epsfig}
\usepackage{amsmath}
\usepackage{bm}

\newcommand{\beq}{\begin{equation}}
\newcommand{\eeq}{\end{equation}}
\newcommand{\bea}{\begin{eqnarray}}
\newcommand{\eea}{\end{eqnarray}}
\newcommand{\bei}{\begin{itemize}}
\newcommand{\eei}{\end{itemize}}

\newcommand{\br}{{\mathbf{r}}}

\newcommand{\bz}{{\mathbf{z}}}
\newcommand{\bp}{{\mathbf{p}}}

\newcommand{\etal}{{\em et al.}}

\def\subB{{\mbox{\tiny{B}}}}

\def\suhe#1{{\bf #1}:}

\begin{document}

\title{Pairing in ultracold Fermi gases in the lowest Landau level}

\author{G.~M\"{o}ller$^{1}$, Th.~Jolicoeur$^2$, and N.~Regnault$^3$}

\affiliation{$^1$Cavendish Laboratories, J. J.~Thomson Avenue, Cambridge UK, CB3 0HE}

\affiliation{$^2$Laboratoire de Physique Th\'eorique et Mod\`eles statistiques, Universit\'e Paris-Sud, 91405 Orsay, France}

\affiliation{$^3$Laboratoire Pierre Aigrain, D\'epartement de Physique, ENS, CNRS, 24 rue Lhomond, Paris,  F-75005}

%%%%%%%%%%%%%%%%%%%%%%%%%%%%%%%%%%%%%%%%%%%%%%%%%%%%%%%%%%%%%%%%%%%%%%%%%%%%%%%%%%%%%%%%%%%%%%%%%%%%%%%%%
\begin{abstract}
We study a rapidly rotating gas of unpolarized spin-$1/2$ ultracold fermions in the two-dimensional regime
when all atoms reside in the lowest Landau level. Due to the presence of the spin degree of freedom both
 $s$-wave and $p$-wave interactions are allowed at ultralow temperatures.
We investigate the phase diagram of this system as a function of the filling factor 
in the lowest Landau level and in terms of the ratio between $s$- and $p$-wave interaction strengths. 
We show that the presence of attractive interactions induces a wide regime of phase separation with formation of 
maximally compact droplets that are either fully polarized or composed of spin-singlets.
In the regime with no phase separation, we give evidence for fractional quantum Hall states. 
Most notably, we find two distinct singlet states at the filling $\nu =2/3$ for different interactions. 
One of these states is accounted for by the composite fermion theory,
while the other one is a paired state for which we identify two competing descriptions with
different topological structure. This paired state may be an Abelian liquid
of composite spin-singlet Bose molecules with Laughlin correlations.
Alternatively, it may be a known non-Abelian paired state, indicated by good overlaps with the
corresponding trial wavefunction.
By fine tuning of the scattering lengths it is possible to create the 
non-Abelian critical Haldane-Rezayi state for $\nu =1/2$
and  the permanent state of Moore and Read for $\nu =1$. For purely repulsive
interactions, we also find evidence for a gapped Halperin state at $\nu=2/5$.
\end{abstract}
%%%%%%%%%%%%%%%%%%%%%%%%%%%%%%%%%%%%%%%%%%%%%%%%%%%%%%%%%%%%%%%%%%%%%%%%%%%%%%%%%%%%%%%%%%%%%%%%%%%%%%%%%%
\date{July 7, 2008}

\maketitle

\section{Introduction}

Cold atomic gases are an ideal system for the study of novel quantum phenomena,
due to the increasingly versatile experimental techniques available to control these systems.
Recently an intense activity focused onto the case of spin-1/2 fermions with attractive interactions.
In the realm of ultralow temperatures, the dominant process of interaction is 
$s$-wave scattering which is allowed by the Pauli principle
when colliding fermions are in a spin singlet state.
The strength of this scattering can be tuned through
Feshbach resonances by applying a static external magnetic field.
On one side of the resonance there is no real bound state close to zero energy
while on the other side there are weakly bound diatomic molecules. If we now consider
a gas of atoms then the side with no bound state
will lead to a BCS instability with formation
of a ground state with pairing correlations while the other side will lead to
Bose condensation of molecules. The crossover between these two regimes
has been the subject of many theoretical condensed-matter studies -- atomic physics 
experiments are now able to probe this important regime.
Superfluidity can be observed in this system of fermions
by imposing a rotation to the gas. There is then occurrence of quantized vortices
arranged in a regular lattice.\cite{Zwierlein}  
This is an interesting  parallel with atomic Bose gases which also display
the Abrikosov lattice of vortices when set in rotation.

In the case of atomic Bose gases it is expected that if the rotation is fast enough,
and the confinement along the rotation axis is sufficiently strong, 
the gas will flatten and reach a two-dimensional regime. There is then formation of
Landau energy levels familiar from the quantum mechanics of a particle in a magnetic field.
When all bosons reside in the lowest Landau level (LLL) there is formation of 
fractional quantum Hall (FQH) states of the bosons after melting of the vortex 
lattice.\cite{WG00,CWG01,RJ03,RJ04} 
It is thus a very natural question to investigate what happens under
similar circumstances to a gas of spin-1/2 fermions.
Rotation of the system translates into two additional forces in the rotating frame~:
the centrifugal force and the Coriolis force. In a trap with a harmonic
confining potential the restoring force is linear in distance from the axis
of rotation, as is the centrifugal force. For fast enough rotation these forces may 
thus nearly compensate each other and we are left with the Coriolis force
which is formally equivalent to a static external magnetic field applied
on the neutral atoms. Under these conditions, the two-dimensional regime is very special 
since the kinetic energy is quenched and one-particle energy levels form
degenerate sets -- the Landau levels responsible for the appearance of the
quantum Hall effect in condensed matter physics. At low enough temperatures
all particles will occupy the LLL and solely the interactions between them
will determine the nature of the ground state~: there is no longer
any competition between kinetic energy and potential interaction energy.
When there are only few bosons per available quantum state in the LLL,
correlated liquid phases with special properties emerge, the so-called
fractional quantum Hall liquids. For spinless bosons interacting
via $s$-wave scattering, the filling factor 1/2 leads to a ground state
which is exactly given by the Laughlin wavefunction.\cite{WG00}  At other filling factors
there are other  FQH states, some that belong to the standard lore~\cite{RJ03}
and some more exotic states.\cite{CWG01} 
If we now consider ultracold spin-1/2 fermions, which may, in principle,
reach the same quantum Hall regime. The Zeeman energy that lifts the degeneracy between the 
two fermion species can be manipulated but in this paper we focus specifically on the
case of zero Zeeman energy. In the context of the quantum Hall physics of electrons,
this case can be realized only under very special circumstances such as large
external hydrostatic pressure applied to the sample. It is known that even with complete
spin degeneracy, the electrons prefer to adopt a fully polarized ground state
at least for fractions like $\nu =1$ and $\nu =1/3$. Ultracold spin-1/2 fermions 
have interactions that are of a very different nature compared to electrons
in semiconductor heterostructures.
Scattering of spin-1/2 fermions at ultralow temperatures is normally dominated by
$s$-wave processes and $p$-wave scattering is suppressed. However manipulation
of Feshbach resonances can be used to boost $p$-wave scattering up to the same order
of magnitude of $s$-wave interactions in
cold gases of potassium~\cite{Regal03,Ticknor04,Guenter05} as well as 
lithium.\cite{Zhang04,Schunck05,Chevy05}  It is thus physically relevant to explore
the physics as a function of the ratio of the $s$- and $p$-wave scattering lengths.

In this paper, motivated by experimental advances, we investigate the quantum Hall physics
of spin-1/2 ultracold fermions. We concentrate on the balanced case with equal
populations of both spin states.  In the LLL there are now two relevant parameters:
the filling factor and the ratio of $s$-wave and $p$-wave scattering. We stress that only the
ratio is relevant since in the absence of kinetic energy the overall energy scale
factors out of the physics. The complete disappearance of kinetic energy in the LLL
leads to a wide regime of phase separation if there is attraction between the atoms
in some allowed spin channel. The atomic system then prefers to form a maximally compact
droplet of spin zero if there is $s$-wave attraction and a ferromagnetic droplet with maximal spin
when there is attraction in the $p$-wave channel. These states have a very simple explicit form 
but they do not exhaust all the possibilities for the ground states of the system.
To characterize the quantum Hall states we use exact diagonalizations of the many-body problem
in the spherical geometry. Candidate quantum Hall liquids have a very definite
ratio of flux vs number of particles which depends upon their internal topological order
and can be used as an identifying signature. 
We find evidence for incompressible quantum Hall states at the filling fraction
$\nu =2/3$. If the $s$-wave interaction is repulsive enough there is formation
of a singlet state which can be described by standard composite fermion construction.
For weaker interactions in the $s$-wave channel there is a transition towards a state
which we tentatively describe as an Abelian paired state as envisioned by Halperin~\cite{Halperin83}
with singlet molecules forming a standard Bose-Laughlin state at an effective filling factor
$\nu_\subB =1/6$. 
However this may not be the whole story since we also find  very good overlap
with a non-Abelian paired state introduced by Ardonne et al.~\cite{Ardonne02}.
At $\nu=2/5$, we identified a gapped spin-singlet state described
by a Halperin wavefunction.\cite{Halperin83}
%\marp{check statement about $\nu=2/5$!}
Finally we also show that there are critical points, i.e. gapless systems,
at filling fractions $\nu=1$, $1/2$ that are described by the 
non-Abelian Haldane-Rezayi state~\cite{HR88} for $\nu =1/2$ 
and the permanent state~\cite{MooreRead,ReadMoore,RR96}
for $\nu =1$. 

In section II we discuss the peculiarities of ultracold fermions with spin in the LLL
in rotating systems. Our results for the various filling factors 
are exposed in section III. Conclusions are given in section IV.

\section{Interacting ultracold fermions with spin in the LLL}

We first discuss the one-body problem for a particle trapped in an anisotropic
rotating potential. We consider the case when there is strong confinement along 
the z-axis which is also the rotation axis.
The one-body Hamiltonian in the rotating frame can be written as:
\begin{equation}
\mathcal{H}_R= \frac{1}{2M}\left[\bp -M\Omega \hat \bz \times \br\right]^2
+ \frac{1}{2} M \omega_z^2 z^2
+ \frac{1}{2} M (\omega_\perp^2-\Omega^2) \br_\perp^2 .
\end{equation}
In this equation, $M$ is the mass of the fermion, $\Omega$ is the rotation velocity, 
$\omega_z$ is the characteristic trapping frequency along the $z$-axis, $\omega_{\perp}$
is the  trapping frequency in the $x$-$y$-plane perpendicular to $z$, and
the coordinates in the $x$-$y$-plane are $\br_\perp^2 =x^2+y^2$.
We assume that the dynamics of the system is effectively two-dimensional,
with the motion along the $z$-direction confined to the lowest eigenstate of the harmonic confinement
$\phi_z \propto \exp[-z^2/2\ell_z^2]$, with $\ell_z=\sqrt{\hbar/M\omega_z}$.
The quantum Hall regime may be recovered when $\Omega\approx\omega_{\perp}$.
The Coriolis force, which is formally equivalent to the Lorentz force, mimics a magnetic
field $B = 2M\Omega$. The one-body eigenstates are then given by the two-dimensional
Landau levels: they are a set of highly degenerate states with a spacing given by the cyclotron
frequency $\hbar\omega_c=2\hbar\Omega$, their degeneracy being proportional to the area of the system.
The eigenfunctions of the LLL are given by:
\begin{equation}
\phi_{m}(z) =\dfrac{1}{\sqrt{2^{m+1}\pi m!}} \,z^m \,{\rm e}^{-|z|^2/4\ell_0^2},
\end{equation} 
where $z$ is the complex coordinate in the $x$-$y$-plane, the length scale
is set by the ``magnetic'' length  $\ell_0 = \sqrt{\hbar/2M\Omega}$
and $m$ is a \textit{positive} integer.
If we are not exactly at the critical rotation velocity there will be a remaining harmonic
potential. This residual effect is not expected to affect states that are incompressible 
i.e. robust to density changes due to a bulk gap. This is the case of the so-called
fractional quantum Hall states that we study in this paper.

In this work we focus on the case of two species of fermions. This is the situation that
is relevant to the study of the BEC-BCS crossover. We will consider two degenerate species
and treat them as spin-1/2 fermions. Here degenerate means
that the energy splitting between these states is much smaller than their interaction energy.
The ``spin'' that we use may have a complex microscopic origin. For example,
in $^6$Li in zero magnetic field the low lying states are a hyperfine doublet
$F=1/2$ and a quartet $F=3/2$. With a moderate field the ground state becomes a triplet
and the two low-lying states of this triplet are of particular interest
to create stable spin mixtures. Notably it is known that there is a pronounced Feshbach
$s$-wave resonance in the scattering between these two states.
In addition there is a also a $p$-wave resonance between these states.
By tuning the magnetic field it is thus possible to have some control
of the relative scattering strength between $s$-wave and $p$-wave interactions.
We now show how one can parametrize the interaction Hamiltonian in the LLL.
It is well known that the two-body problem in the LLL is trivially solvable
for arbitrary rotationally invariant interactions and that the eigenenergies are given
by the set of so-called Haldane pseudopotentials~\cite{Haldane83} $V_m$:
\begin{equation}
 V_m =\dfrac{\langle\phi_m|V(z)|\phi_m\rangle}{\langle\phi_m|\phi_m\rangle},\quad m\geq 0,
\end{equation} 
where $V(z)$ is the two-dimensional interaction potential and 
$\phi(z)\propto z^m\exp (-|z|^2/8\ell_0^2)$ is the relative particle eigenstate
with angular momentum $m$. The even values of $m$ are relevant to the spin singlet
channel and the odd values to the spin triplet channel. The $s$-wave (resp.~$p$-wave) scattering 
amplitude is related to $V_0$ (resp.~$V_1$). The interacting N-body problem is then fully
defined by the Hamiltonian:
\begin{equation}
\mathcal{H}=\sum_{m\geq 0}V_m\sum_{i<j} \mathcal{P}_{ij}^{(m)},
\end{equation} 
where $\mathcal{P}_{ij}^{(m)}$ projects the pair of particles $i$ and $j$
onto relative angular momentum $m$. Neglecting scattering in higher partial waves,
the problem only involves $V_0$ and $V_1$ given by:
\begin{equation}
 V_0 =\sqrt{\frac{2}{\pi}}\dfrac{\hbar^2}{2M}\frac{a_s}{\ell_0^2 \ell_z},
\quad
V_1 =\sqrt{\frac{2}{\pi}}\dfrac{\hbar^2}{M}\frac{a_p^3}{\ell_0^4 \ell_z}.
\end{equation} 
Here $a_s$ (resp.~$a_p$) is the $s$-wave (resp.~$p$-wave) scattering length
deduced from low-energy scattering limit and we have used explicitly the fact
that the wavefunction along $z$ is the Gaussian ground state wavefunction
of width $\ell_z$. 
Note there is no explicit spin dependence in the interactions, it is only
 through the Pauli principle that the spin degrees of freedom 
are interacting non-trivially.
Since one can factor out an overall energy scale
the N-body problem in the LLL is thus function only of the ratio $V_1/V_0$.
It is then convenient to parametrize the interaction Hamiltonian as:
\begin{equation}
\label{Htheta}
\mathcal{H}_\theta = g_0\cos\theta\sum_{i<j} \mathcal{P}_{ij}^{(0)}
  +g_0\sin\theta\sum_{i<j}\mathcal{P}_{ij}^{(1)},
\end{equation} 
where $g_0$ is the overall energy scale.
The phase diagram can be represented as a circle described by the angular variable
$\theta$.

Our strategy is to perform exact diagonalizations of Eq.~(\ref{Htheta})
for a small number of fermions in the spherical geometry. This kind of
calculations pioneered by Haldane has proved fruitful to find
incompressible quantum Hall states.
Technical details can be found in the work of Fano \etal \cite{FanoOrtolani} 
It is convenient to switch to the equivalent magnetic language
in which vorticity is now magnetic flux. A sphere can be pierced only by
an integer number of flux quanta $N_\phi=2S$ and the Landau levels
can be classified according to their orbital angular momentum.
The LLL has momentum $S$, and hence is $2S+1$ times degenerate and is spanned
by functions $u^{S+M}v^{S-M}$,
$u =\cos (\theta/2) {\rm e}^{-i\phi/2}$ and $v=\sin (\theta/2) {\rm e}^{+i\phi/2}$,
where $M=-S\dots +S$. There are again pseudopotential
parameters as in the infinite plane described above and only two of them are relevant
in our case:
\begin{equation}
 V_0 =\frac{(2S+1)^{2}}{S(4S+1)}\sqrt{\frac{2}{\pi}}\dfrac{\hbar^2}{2M}\frac{a_s}{\ell_0^2 \ell_z},
%\equiv \frac{(2S+1)^{2}}{S(4S+1)} \frac{\ell_0'^2}{\ell_0^2} g_s,
\quad
 V_1 =\frac{(2S+1)^{2}}{S(4S-1)}\sqrt{\frac{2}{\pi}}\dfrac{\hbar^2}{M}\frac{a_p^3}{\ell_0^4 \ell_z}.
%\equiv \frac{(2S+1)^{2}}{S(4S-1)} \frac{\ell_0'^4}{\ell_0^4} g_p.
\end{equation} 
We consider the family of Hamiltonians $\mathcal{H}_\theta$ on the sphere, allowing a single
parameter $\theta$.
Many-body states on the sphere are classified by their total angular momentum $L$ and their spin $S$.
In order to search for incompressible quantum states we diagonalize the Hamiltonians $\mathcal{H}_\theta$ 
in the LLL for finite systems with even numbers of particles $N=2N_\uparrow = 2N_\downarrow$, i.e., total 
spin $S_z=0$, at different flux $N_\phi$. Candidates for incompressible states have full rotational 
invariance on the sphere and can be distinguished by their angular momentum $L^2=0$.
Such states form families defined by a specific relationship between flux and number of particles
of the form $2S=(1/\nu)N-\sigma$ where the constant $\sigma$ is called the \textit{shift}.
This shift -- which is irrelevant in the thermodynamic limit -- is nevertheless
a very useful tool to differentiate between states having different internal structures.

We adopt the convention to measure energies in terms of density corrected magnetic lengths 
$\ell_0'= \ell_0 \sqrt{2S\nu/N}$.\cite{MorfAmbrumenil86} Note that this leads to different 
finite size scalings for the $s$- and $p$-wave channel.
We separate the factors determining the 
finite-size scaling from the coupling constants $g_s$ and $g_p$ setting the scale of interactions 
in the two scattering channels we consider:
\begin{equation}
 V_0 \equiv \frac{(2S+1)^{2}}{S(4S+1)} \frac{\ell_0'^2}{\ell_0^2} g_s, \quad
 V_1 \equiv \frac{(2S+1)^{2}}{S(4S-1)} \frac{\ell_0'^4}{\ell_0^4} g_p.
\end{equation} 
At a given number of particles $N$ and filling factor $\nu$, we then define the interaction 
Hamiltonian $\mathcal{H}_\theta$ by setting $g_s=\cos(\theta)g_0$ and $g_p=\sin(\theta)g_0$,
thus defining the overall energy scale $g_0$.

\section{Phase separation}
\label{phase-sep}

\begin{figure}[ttbb]
\begin{center}
    \includegraphics[width=10.0cm]{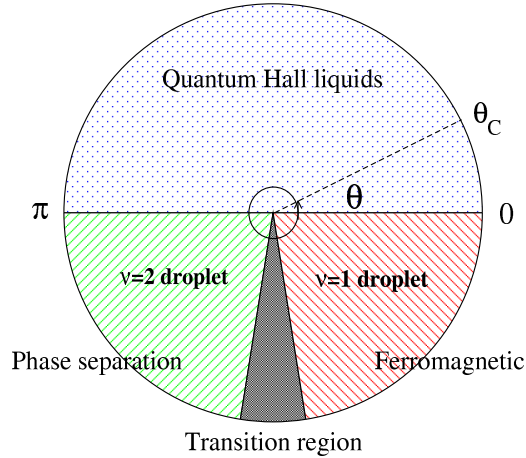}
  \end{center}
\caption{\label{fig:regimes} (color online)
Schematic phase diagram for balanced Fermi gases in the LLL: 
For $0\lesssim\theta\lesssim\pi$, or $V_1>0$, several
incompressible quantum liquids can be realized. 
For $V_1<0$, i.e., $-\pi \lesssim \theta \lesssim 0$, 
the sign of $V_0$ determines whether the system is unstable to phase separation into a 
locally $\nu=2$ droplet ($V_0<0$, $-\pi \lesssim \theta \lesssim -\pi/2$) 
or into a ferromagnetic state with locally $\nu=1$ 
($V_0>0$, $-\pi /2 \lesssim \theta \lesssim 0$). The dotted line locates
the value $\theta_C\approx \pi/7$ in the phase diagram.}
\end{figure}

We start our discussion by some general remarks about the phase diagram as a function of $\theta$.
The quantum Hall regime in solid state physics, i.e., electrons interacting through
the Coulomb potential corresponds to all $V_m$ positive and decreasing with $m$.
This is realized in the upper right quadrant of the circle in Fig.~\ref{fig:regimes}.
We expect to recover the physics of electronic FQHE with zero Zeeman energy there.
Indeed the ratio $V_1/V_0$ equals 1/2 for pure Coulomb interaction in the LLL
(neglecting the influence of layer width)
and this means $\theta = {\rm atan}\, (1/2)\approx 0.46$ close to $\pi /7$.

The possibility of attractive channels may lead to novel features in the remainder of the phase
diagram. In the upper left quadrant we have attraction in the $s$-wave spin-singlet channel 
competing with repulsive $p$-wave interactions: hence there will be an interplay between pairing 
and Laughlin-like correlations.

\begin{figure}[thpp] 
  \begin{center}
    \includegraphics[width=8.0cm]{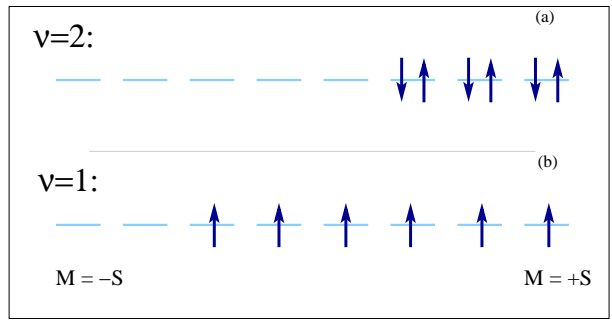}
  \end{center}
  \caption{\label{droplets} (color online)
The maximum density droplets that can form in a system with attractive interactions.
When there is attraction in both s and p wave channels it is best to make a spin singlet droplet
of maximum density. This is constructed by filling exactly all orbitals with a singlet pair (a).
When there is attraction in the $p$-wave channel but repulsion in the $s$-wave channel
the maximum density droplet is now fully spin-polarized, i.e. it is a local $\nu= 1$ 
quantum Hall ferromagnet (b). The state at $S_z=0$ as studied here is the spin rotation of
state (b) into the $x$-$y$-plane.}
\end{figure}

If both $s$- and $p$-wave interactions are attractive, as is the case for
$\pi\lesssim\theta\lesssim3\pi/2$, one has to consider the possibility
of phase separation. Indeed in the LLL the kinetic energy is frozen and is
not an obstacle to clustering of particles in a state of maximal density
favored by attractive interactions. In the lower left quadrant
it is thus favorable to make S=0 pairs to maximize the $V_0$ interaction
and then to put all pairs as close as possible to maximize $V_1$. The state
is then simply a Slater determinant constructed by occupation of orbitals 
of the LLL by singlet pairs. This state is a droplet with maximum local density
and its local filling factor is thus $\nu =2$. 
The occupation pattern is described in Fig.~\ref{droplets}.
This phenomenon happens for any number of particles and is always clearly observed
in our numerical studies at all filling factors. The ground state is then a state with the 
maximal possible angular momentum $L_\text{tot}$. The value is
given by the maximum possible projection of the angular momentum
$L_\text{tot}=L_z^\text{max}$ where
$L_z^\text{max}=2S+2(S-1)+2(S-2)+...+2(S+1-N_e/2)$.

Finally, for $3\pi/2\lesssim\theta\lesssim 2\pi$, there is also phase separation but now
with formation of a fully ferromagnetic droplet as pictured in Fig.~\ref{droplets}.
The local filling factor is now $\nu =1$. The maximum value of the momentum
is obtained by occupying each orbital exactly once:
$L_\text{tot}=L_z^\text{max}$ where
$L_z^\text{max}=S+(S-1)+(S-2)+...+(S+1-N_e)$.

In contrast to incarnations 
of this state in electronic bilayer systems, there is no charging energy for unbalanced spin-populations in cold 
atomic gases since these systems are neutral. We conclude that this phase has a strong susceptibility to rotations
of the total spin out of the $x$-$y$-plane. This ferromagnetic phase supports skyrmion physics 
which provides opportunities 
for novel experimental probes of this phenomenology in atomic gases. In our finite system simulations, 
we find that the transition between the $\nu=2$ and $\nu=1$ regimes is not direct, but proceeds via 
intermediate states. 
We note also that these two regimes were also found to arise naturally as the limiting two-dimensional 
behavior of (spin-) density waves of rapidly rotating Fermi gases with $s$-wave interactions in a
three-dimensional regime for $V_0>0(<0)$.\cite{MC07}

\section{Incompressible states}

Due to the limitation of numerical calculations to a small number of particles, we concentrate on the
most prominent filling fractions. With the presence of attraction in the $s$-wave channel,
it is expected that some kind of pairing will play a role. Wavefunctions including pairing
have been suggested by Halperin~\cite{Halperin83} in the context of electronic systems.
If two fermions with spin-1/2 form a bound singlet pair, then this fluid of pair will have charge two
and thus feels a flux which is twice the original flux. 
The number of pairs is also half the number of fermions.
Now these pairs can have Laughlin-like correlations under appropriate
circumstances. They will have then a special flux-number of particle relationship:
\begin{equation}
\label{boseshift}
 2\times 2S =m_\subB\left(\dfrac{N}{2}-1\right),
\end{equation} 
where pairs form a Laughlin fluid with exponent $m_\subB = 2,4,...$ with a Bose filling factor
$\nu_\subB=1/m_\subB$
The original fermions then form a state with $\nu = 4/m_\subB$ which is a spin-singlet.
The spectral signature of such a state is a singlet ground state and it is natural to expect
that the excited states display a well-defined magnetoroton branch resulting from
the Laughlin nature of the Bose fluid. This branch should thus extend up to angular momentum
$L=N/2$, i.e., to the number of (composite) \textit{bosons}. Such Abelian paired state are not 
thought to be realized in electronic systems
even at zero Zeeman energy. We expect that FQHE states describable by a composite fermion
construction are relevant to small values of $\theta$ close to the Coulomb value
$\theta_C \approx \pi/7$.

In the following section, we discuss the different quantum Hall states that are realized for 
repulsive $p$-wave interactions at various filling factors $\nu$. 
We have searched for systems with a singlet ground state
as a function of the number of particles and flux. Due to the exponential growth of 
the size of the Hilbert space, only a small set of values can be investigated in detail.
Strictly speaking, identification of a fractional quantum Hall state requires
finding a whole series of states with a definite relation between number of particles and flux
with a smooth behavior of physical observables with system size increasing towards the thermodynamic limit.
Practically, one should keep in mind that our assignments to quantum Hall fractions are tentative.
We are guided by the previously mentioned scheme of Abelian pairing (\ref{boseshift})
and by the standard composite fermion construction.

%%%%%%%%%%%%%%%%%%%%%%%%%%%%%%%%%%%%%%%%%%%%%%%%%%%%%%%%%%%%%%%%%%%%%%%%%%%%%%%%%%%%%%%%%%%%%%5
\subsection[Filling factor nu = 2/3]{Filling factor $\nu = 2/3$}

We first discuss the possibility of observing the Abelian paired state for $m_\subB=6$.
The reasoning above leads to a filling factor $2/3$ and a shift equal to $-3$. We find
strong candidate states for $N=6$, $8$, $10$ in wide range of values of $\theta$ that includes
the so-called hollow-core point $\theta =\pi/2$. In all cases there is a clear gap between
a $L=0$, $S=0$ ground state and excited states, the lowest-lying excited states having
the expected structure of a magnetoroton branch extending up to $L=N/2$ and having $S=0$,
i.e., they are made of excitations of unbroken singlet pairs. This is most clearly seen 
in the center of the gapped phase around $\theta\approx \pi/2$.
However since there is no available explicit simple wavefunction for the Abelian paired
state we cannot perform any overlap calculation.
For large values of $\theta$ these states are killed by a collapse towards phase separated 
maximum density droplets. The situation is more interesting for small values  of $\theta$. 
Here we find a transition with a collapse of the gap for $\theta \approx 0.2\,\pi$ beyond which 
excited states are nearly degenerate. 
%\marp{didn't understand what you meant by 'the states become degenerate'}
This transition appears as a rapid crossover as a function of $\theta$.
For smaller values of $\theta$ we instead observe another series of states with a shift equal to $-1$,
and this is exactly what we expect from the composite fermion construction which known 
to be relevant for Coulomb interactions. In the realm of electrons there is a quantum Hall state
at $\nu =2/3$ which is fully spin-polarized and is simply the particle-hole conjugate
of the celebrated Laughlin state at $\nu =1/3$. However it has been observed~\cite{E1,E2,Engel}
that reduction of the Zeeman energy leads to the disappearance of this state
and formation of a spin-singlet state with the same filling. This state can be explained 
straightforwardly in the composite fermion construction with spin.\cite{WDJ93,WJ94}  
In this scheme the CFs exactly fill
the effective lowest Landau level with one singlet pair for each available orbital.
One has thus $2(|2S^*|+1)=N$ where $2S^*$ is the reduced flux felt by the CFs. To reach
total filling $\nu =2/3$ this effective flux is negative and given by $2S^*=2S-2(N-1)$. Hence
this series of S=0 states has $2S=(3/2)N-1$. For the case of electrons with 
Coulomb interactions and zero Zeeman energy, it has been shown that CF wavefunctions
constructed from this series have extremely good overlap with the numerically obtained ground states
(the overlap is 0.998 for N=8 and 0.99896 for N=6 from ref.~\onlinecite{WDJ93}).
We have computed the overlap between the Coulomb ground state and the ground state
of the family of Hamiltonians indexed by $\theta$ and found that the overlap rises up to
0.9996 for $\theta = 0.025\pi$. It is extremely close to unity in the region where $V_0$
is larger than $V_1$: this means that the physics can be described by the composite fermion
scheme.

Concerning the state at shift -3, we note that there is another interesting candidate~\cite{Ardonne02}
which is a paired state and also a spin-singlet. Its wavefunction is
given by~:
\begin{equation}
\label{eq:non_abelian_2_3}
\Psi_\text{NASS} = \mathrm{Pf}\left(\dfrac{1}{z_i-z_j}\right)
\prod_{i<j}({z_{i}^{\uparrow}-z_{j}^{\uparrow}})^2
\prod_{i<j}({z_{i}^{\downarrow}-z_{j}^{\downarrow}})^2
\prod_{i, j}(z_i^{\uparrow}-z_j^{\downarrow}).
\end{equation} 
In this equation  the $z_i$'s stand for all particle coordinates and $z_i^\uparrow,z_i^\downarrow$
are the coordinates of the two spin components (this notation is a shorthand for the 
full wavefunction with both spin and space coordinates, it is standard technology to reconstruct
the complete state from this notation).
The Pfaffian of an antisymmetric matrix is defined as $\mathrm{Pf}(A_{ij})={\mathcal A}[A_{12}A_{34} \dots]$ 
with $\mathcal A$ denoting antisymmetrization. 

\begin{figure}[ttbb]
\begin{center}
  \includegraphics[width=0.7\columnwidth]{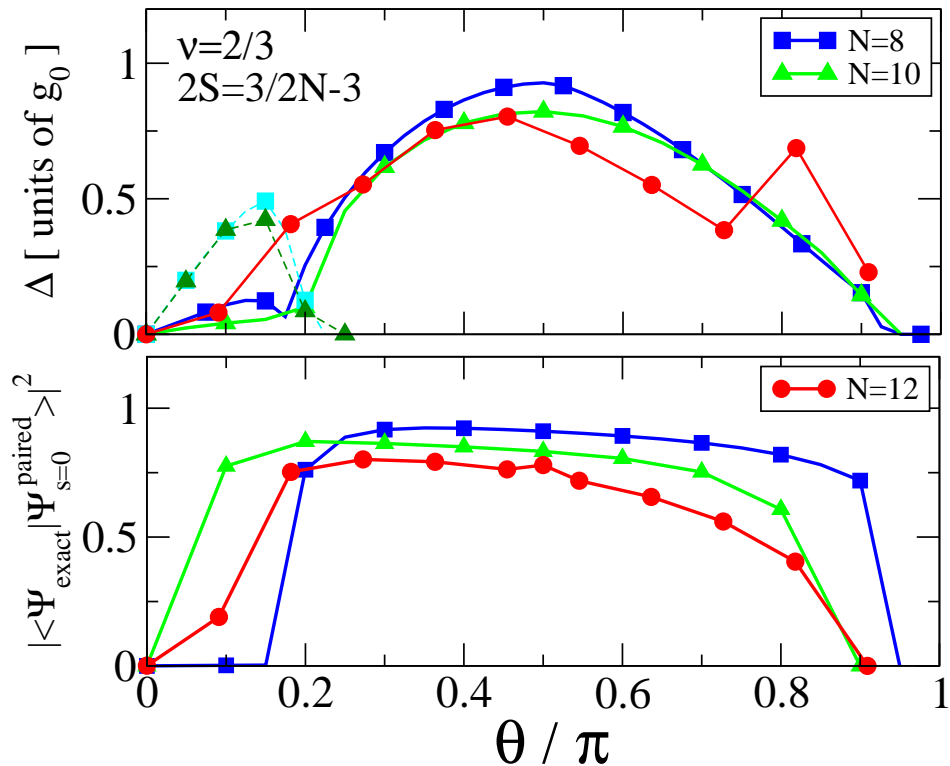}
\end{center}
\caption{\label{fig_2_3} (color online)
Top panel~:
excitation gaps between the ground state and the first excited state
as a function of parameter $\theta$ for sizes $N=8$, $10$. The solid lines
are computed for the flux $2S=(3/2)N-3$ as expected for an Abelian paired state.
with filling factor $\nu=2/3$. This state is destroyed for $\theta \lesssim 0.2\,\pi$
and replaced by the composite fermion state with $S=0$ and filling $2/3$ with shift
$2S=(3/2)N-1$ as shown with dashed lines. Energies are measured in terms of the
overall energy scale $g_0$.
Lower panel~:
the overlap between the exact ground state at shift $-3$ and the non-Abelian state constructed
from the Halperin (221) state times a Pfaffian factor. The agreement is very good within
the gapped phase 
}
\end{figure}

The state (\ref{eq:non_abelian_2_3}) state is known to exhibit spin-charge separation~:
the fundamental excitations are spinons with charge zero and spin 1/2 and spinless holons
with charge $\pm 1/3$. The braid statistics of these quasiparticles is non-Abelian;
its properties have been investigated by Ardonne et al.~\cite{Ardonne02}.
The non-Abelian spin-singlet state $\Psi_\text{NASS}$ can be constructed as the unique groundstate 
of an appropriate Hamiltonian composed of (2, 3 and 4-body) hardcore interactions 
and the total spin $S^2$. [This Hamiltonian derives from the corresponding bosonic
state $\Psi_b=\Psi_\text{NASS}/\Psi_1$, which is obtained as the groundstate of a simple 
three-body contact interaction with an additional $S^2$-term (with $\Psi_1$, the wavefunction
for a filled Landau level).]
We have computed the overlap between the candidate state obtained in this manner and the
exact groundstate of our model Hamiltonians. 
Our results are displayed in the lower panel of Fig.~\ref{fig_2_3}, for up to N=12 electrons.
The overlap is extremely high in the regime where we observe evidence for a gapped state.
For N=14 at the point $\theta =\pi/2$ the overlap is still as large as 0.71398.
It is not clear yet if this non-Abelian state has the same excitation structure
that we guessed for Abelian pairing. 
% However present numerical evidence favors its existence in a large range of parameters.

\begin{figure}[ttbb]
\begin{center}
  \includegraphics[width=0.6\columnwidth]{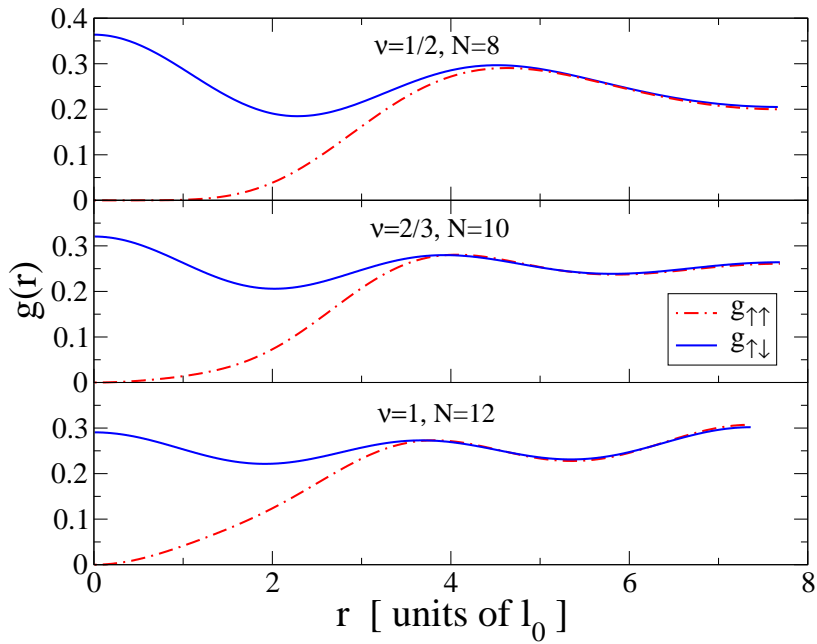}
\end{center}
\caption{  \label{G_up_down} (color online)
Correlation functions as a function of particle separation in units of the magnetic length
for the spin-singlet ground states of the hollow-core Hamiltonian $\theta =\pi/2$ 
(only $p$-wave scattering). From top to bottom: $\nu=1/2$ (N=8), $\nu=2/3$ (N=10) and 
$\nu=1$ (N=12). Only the $\nu =2/3$ state will correspond to a gapped quantum hall state
in the thermodynamic limit (probably). In all cases
there is a  maximum of $g_{\uparrow\downarrow}$ at the origin which is indicative of the paired
character of the state.}
\end{figure}

The paired character of these states is clear if we look at the correlation function
$g(r)$. The Pauli principle requires that for same spin projection this correlation goes
to zero for zero separation: $g_{\downarrow \downarrow}(r\rightarrow 0)\rightarrow 0 $
and $g_{\uparrow \uparrow}(r\rightarrow 0)\rightarrow 0 $. However it does not require
that the opposite-spin correlations go to zero. In composite fermion states it is known
that $g_{\uparrow\downarrow}$ is very small albeit not zero at $r=0$. On the contrary
when we weaken $V_0$ we observe a large increase of $g_{\uparrow\downarrow}(0)$
as was first noted by Haldane and Rezayi~\cite{HR88}: see Fig.~\ref{G_up_down}.
In this context, we also note that pairing of ultracold fermions was found to be relevant
for fermions at filling factor $\nu=2$ interacting only in the $s$-wave channel: this system
supports a paired state of charge-2 bosons.\cite{Yang07}

Finally we note that there was a related numerical study~\cite{McDonald96}
of electrons at filling $\nu=2/3$ in the context of bilayer systems for which the layer index
plays the role of a pseudospin. The inter and intralayer Coulomb interactions are
different from our case. While the study~\cite{McDonald96} gives evidence for the $\nu=2/3$
spin-singlet composite fermion state, there is also evidence for stability of a decoupled state
of type (330) in Halperin notation in some part of the phase diagram. Here, in the ultracold
atoms system, we find that this state has always low overlap with the ground state.

\begin{figure}[thpp] 
  \begin{center}
    \includegraphics[width=0.5\columnwidth]{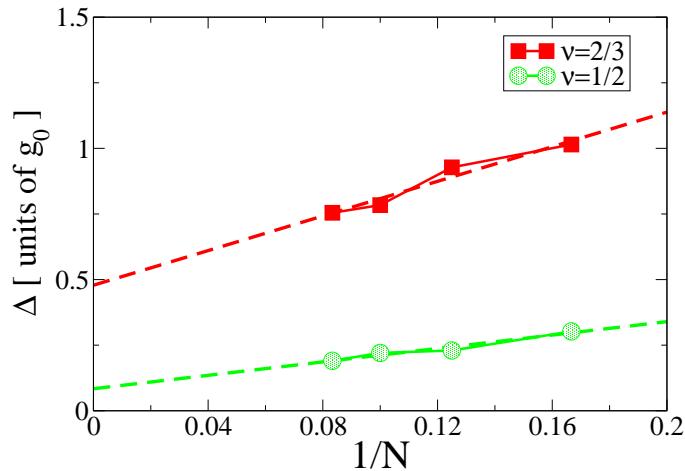}
  \end{center}
\caption{\label{fits}
(color online)
Extrapolation of the gaps for the hollow core model ($\theta =\pi/2$, i.e. with only $p$-wave interactions),
expressed in units of the coupling $g_0$. The asymptotic values from linear extrapolation over $N^{-1}$ are
$\Delta=0.48(6)$ for $\nu=2/3$ and compatible with zero for $\nu=1/2$.
The points for $\nu =2/3$ are taken along the series with shift $\sigma =3$ as expected from an Abelian 
paired state. For $\nu =1/2$ we use the shift of the Haldane-Rezayi state, i.e. $2S=2N-4$, which is the
maximum density droplet with $\langle E \rangle =0$.
This is evidence for an incompressible fluid at $\nu =2/3$ and consistent with the Haldane-Rezayi state
at $\nu =1/2$ being gapless.}
\end{figure}

%%%%%%%%%%%%%%%%%%%%%%%%%%%%%%%%%%%%%%%%%%%%%%%%%%%%%%%%%%%%%%%%%%%%%%%%%%%%%%%%%%%%%%%%%%%%%%%%%%%5
\subsection[Filling factor nu = 1]{Filling factor $\nu = 1$}

\begin{figure}[ttbb]
\begin{center}
  \includegraphics[width=0.7\columnwidth]{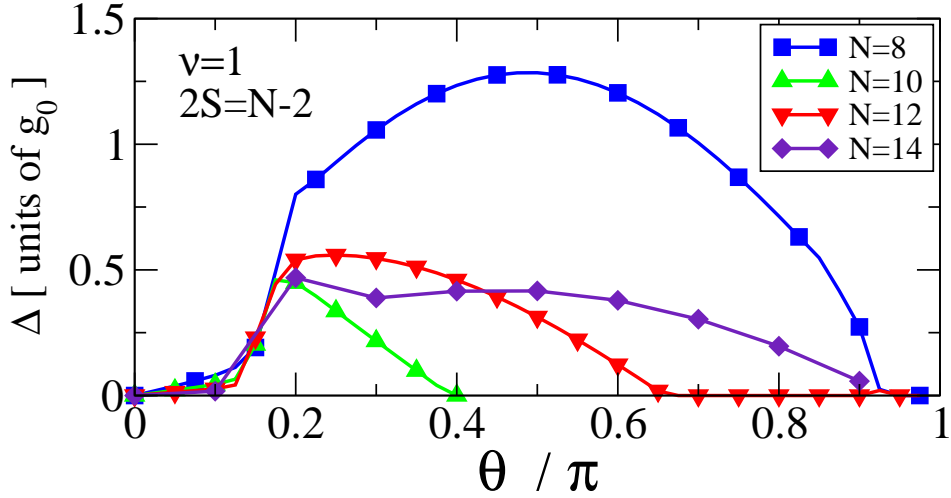}
\end{center}
\caption{  \label{fig_nu_1} (color online)
Neutral excitation gaps of the spin-singlet states 
for $2S=N-2$. While the $N=8$ and $N=14$ data suggests the appearance
of an incompressible state this conclusion is not supported given the aliasing 
with smaller or vanishing gaps at $N=10$ and $N=12$. 
Close to $\theta \approx 0.12\pi$ there is a transition towards the ferromagnetic maximally
compact state -- the groundstate at this transition point is described by the permanent state of Read 
and Rezayi. Just to the right of this point, there is some evidence that suggests that the gap
may survive in the thermodynamic limit.}
\end{figure}

At filling factor unity there is a possibility that the system forms the fully aligned ferromagnetic
state with an occupancy probability of one for all orbitals in the LLL. The wavefunction is then
given by products of Vandermonde determinant factors:
\begin{equation}
 \label{111}
\Psi_{\rm 111}= 
\prod_{i<j}({z_{i}^{\uparrow}-z_{j}^{\uparrow}})
\prod_{i<j}({z_{i}^{\downarrow}-z_{j}^{\downarrow}})
\prod_{i, j}(z_i^{\uparrow}-z_j^{\downarrow}),
\end{equation} 
where we have omitted the overall Gaussian factor. The notation $\Psi_{\rm 111}$
stands for the powers appearing in the various Jastrow factors and was introduced
by Halperin in his work on multicomponent systems.\cite{Halperin83} 
This state, which is the droplet state that always appears in the lowest right-hand part
of the phase diagram, has shift $\sigma=1$ on the sphere. In the zero Zeeman energy limit of the Coulomb problem
it is well-known that adding or removing one quantum of flux leads to the formation
of a skyrmion~\cite{Ed87,Ed91} which has spin zero. This means that there are spin-0
skyrmion states at flux $2S=N-2$ in the neighborhood of $\theta=\theta_C$ corresponding
to the Coulomb problem.

\begin{figure}[ttbb]
\begin{center}
  \includegraphics[width=0.6\columnwidth]{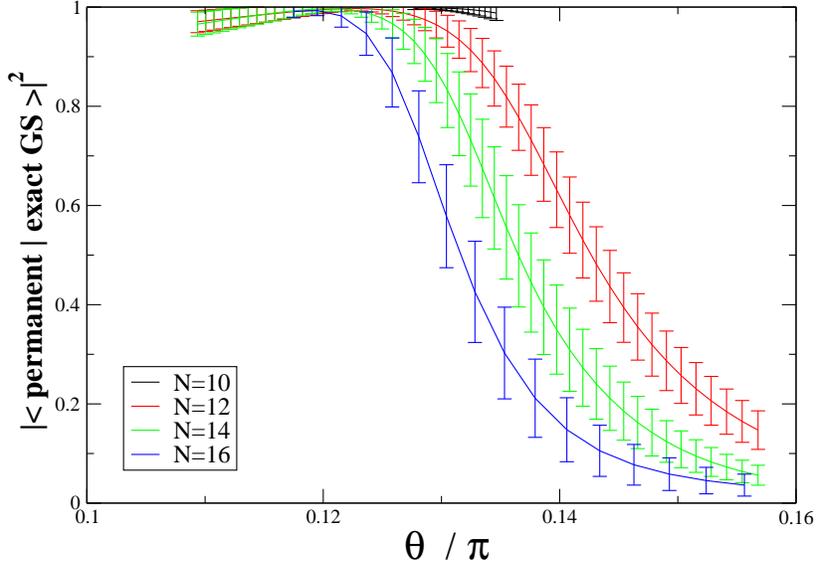}
\end{center}
\caption{  \label{permoverlap} (color online)
The square overlap between the exact ground state for $N=12$, $14$, $16$ and the permanent wavefunction.
The error bars come from the Monte-Carlo process used to evaluate the scalar product.
}
\end{figure}

The Abelian pairing scheme suggests also that there may be a series of incompressible states
for $2S=N-2$ by formation of a Bose Laughlin fluid with $m_\subB=4$. However, we do not consistently
find a non-zero gap over a wide range of values of $\theta$, as in the case of $\nu=2/3$ 
(see Fig.~\ref{fig_nu_1}). While there is
a quite clear gap above a singlet ground state for the systems with $N=8$ and $N=14$, the intermediate
systems with $N=10$ and $N=12$ have a more complicated behavior where the gap drops to zero at 
$\theta \ll \pi$. We interpret these findings as the apparition of a probably compressible state beyond
$\theta\approx 0.17\pi$ which 
replaces the ferromagnetic droplet state mentioned in section \ref{phase-sep}. However, the data
would also be consistent with the existence of an intervening gapped phase in the window 
$0.17\pi\leq \theta\leq 0.35\pi$.
An interesting phenomenon is observed at the point where the transition from the ferromagnetic 
phase occurs~: Read and Rezayi~\cite{RR96} have constructed a wavefunction called the permanent which has
exactly the same shift $\sigma =2$ and is given by:
\begin{equation}
\label{permanent}
\Psi_{\rm perm}= {\rm per}\left(\dfrac{1}{z_{i}^{\uparrow}-z_{j}^{\downarrow}}\right)
\Psi_{\rm 111}.
\end{equation} 
This state is constructed from a conformal field theory
which is non-unitary~\cite{MooreRead,ReadMoore} and is thus expected to be critical.\cite{Read07} 
Read and Rezayi have proposed that this state should occur at the phase boundary of a ferromagnet.
The relationship with the magnetic instabilities has been explored in more detail
in Green's PhD thesis.\cite{GreenThesis} 
We have thus computed the overlap between the permanent state Eq.~(\ref{permanent}) and the ground state
of our model Hamiltonian: see Fig.~\ref{permoverlap} as a function of $\theta$. We find that
right at the transition point seen in the gap, the overlap rises to values extremely 
close to unity: the maximum overlap ranges from $0.998(3)$ for $N=8$ to $0.99(1)$ for $N=14$.
This high overlap we show for the permanent state right where the gap of an adjacent 
ferromagnetic phases collapses gives numerical support for the idea that the permanent state is right 
at the phase termination of a ferromagnet. By analogy with the work of D. Green,\cite{GreenThesis}
it is likely that the phase beyond the permanent has helical order, but our small systems do not allow 
us to analyze this in detail.

%%%%%%%%%%%%%%%%%%%%%%%%%%%%%%%%%%%%%%%%%%%%%%%%%%%%%%%%%%%%%%%%%%%%%%%%%%%%%%%%%%%%%%%%%%%%%%%5
\subsection[Filling factor nu = 1/2]{Filling factor $\nu = 1/2$}

\begin{figure}[ttbb]
\begin{center}
  \includegraphics[width=0.7\columnwidth]{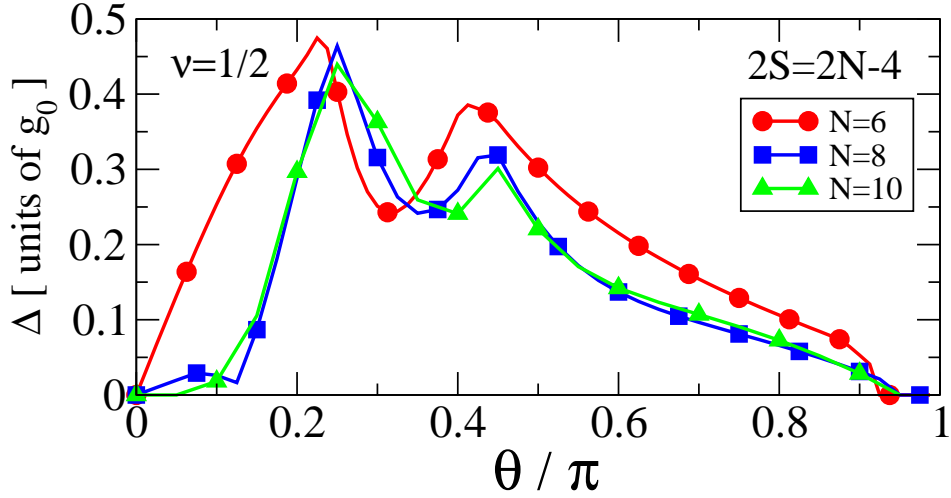}
\end{center}
\caption{ \label{fig_1_2} (color online)
Neutral gaps along the Haldane-Rezayi series $2S=2N-4$. There is no evidence
for a paired incompressible phase.}
\end{figure}

The problem of the nature of Coulomb ground state at $\nu=1/2$ has been of much interest
in electronic systems since there is evidence for an incompressible
quantum Hall state in the \textit{second} Landau level, at $\nu =2+1/2$.
At the present time, the best candidate for describing the FQHE of this state is the non-Abelian 
Moore-Read state.\cite{MooreRead,MS08}  Historically Haldane and Rezayi~\cite{HR88} introduced a candidate
wavefunction which is a spin singlet:
\begin{equation}
 \Psi_{\rm HR}=\det\left(\dfrac{1}{(z_{i}^{\uparrow}-z_{j}^{\downarrow})^2}\right)
\prod_{i<j}(z_i-z_j)^2,
\end{equation} 
where the product in the Jastrow factor runs over all particle indices
irrespective of the spin projection.
The Haldane-Rezayi (HR) state at $\nu=1/2$ occurs for $N_\phi=2N-4$ on the sphere
and is the exact ground state 
of the simplest possible interaction involving both spin species: pure $p$-wave interactions, also known 
as the hollow core model (HCM)~\cite{HR88}: in our language this corresponds to $\theta =\pi/2$.
Since it can be derived from a non-unitary conformal field theory
it is presumably gapless.\cite{ReadGreen,Read07} 
This is what we find form our numerical studies at $\theta=\pi/2$: see Fig.~\ref{fits}.
Extrapolation along this series of states is compatible with zero gap (even though a small
but finite gap cannot be excluded).
The same shift can also be considered as candidate for Abelian paired states
with $m_\subB=8$. However, as in the case of $\nu =1$ we do not find evidence for an extended
gapped phase. Neutral gaps are displayed in Fig.~\ref{fig_1_2}.
The finite-size gap does not peak for pure $p$-wave interactions as for the other filling factors in this family
of states. Rather, there is a two-peak structure which is indicative of the presence
of several phases. The knowledge of the HR critical point right at $\theta=\pi/2$ even suggests
that there may be an extended critical region but our limited data does not allow a firm conclusion.
This is left for future work.

%%%%%%%%%%%%%%%%%%%%%%%%%%%%%%%%%%%%%%%%%%%%%%%%%%%%%%%%%%%%%%%%%%%%%%%%%%%%%%%%%%%%%%%%%%%%%%%5
\subsection[Filling factor nu = 2/5]{Filling factor $\nu = 2/5$}
\begin{figure}[ttbb]
\begin{center}
  \includegraphics[width=0.7\columnwidth]{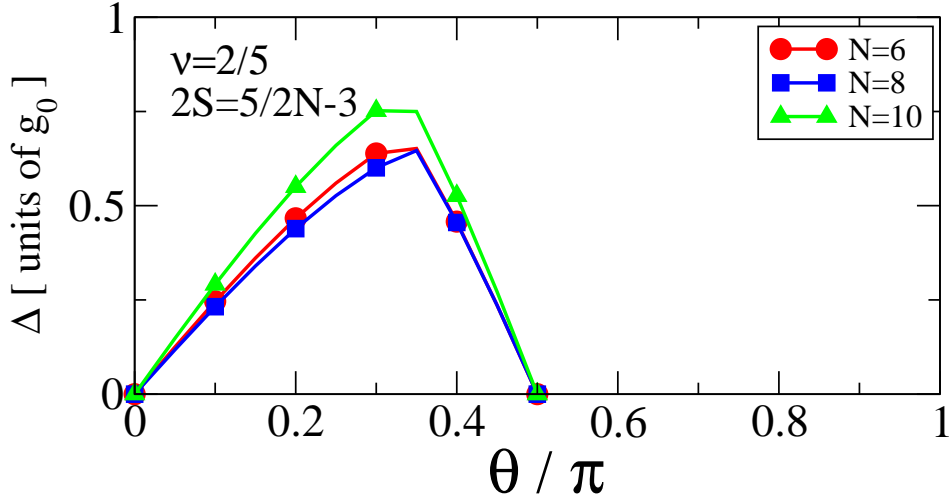}
\end{center}
\caption{\label{fig_2_5} (color online)
The excitation gaps between the ground state and the first excited state
as a function of parameter $\theta$ for sizes $N=6$, $8$, $10$. The solid lines
are computed for the flux $2S=(5/2)N-3$ as expected for the Halperin state $\Psi_{332}$.
with filling factor $\nu=2/5$. This state exists for $ 0 \leq \theta \leq \pi/2$
and its gap goes to zero linearly at either extremity of this interval.
}
\end{figure}

The Hamiltonian $\mathcal{H}_\theta$ yields precisely the Halperin wavefunction 
$\Psi_{332}$ [analogous to Eq.~(\ref{111})] with shift $N_\phi=5/2 N - 3$ as its
exact zero-energy groundstate in the entire sector $0\leq \theta \leq \pi/2$~: For positive pseudopotential
coefficients $V_0$ and $V_1$, a zero-energy state of maximum density is obtained if particles
with same spin have a relative angular momentum $m_\text{rel}\geq 0$ and particles with different 
spin have $m_\text{rel}\geq 1$. The Halperin state $\Psi_{332}$ is the maximum density state
with proper symmetries that satisfies these requirements. The nature of its excitations however 
depends upon $\theta$. For instance, at $N=10$ we find that the lowest excited state has spin $S=2$
for $\theta \lesssim 0.3\pi$ while it becomes a spin singlet at larger $\theta$.
At the HCM point,
the HR state at $\nu=1/2$ instead becomes the maximum density state with $E=0$. From the 
construction of the Halperin state, we expect the gap to be determined approximately by the
smallest value of $V_0$ and $V_1$, with the lowest energy excitation breaking either condition
on the relative angular momentum. Indeed, numerics show a linear behavior of the gap close
to $\theta=0$ and $\theta=\pi/2$~: see Fig.~\ref{fig_2_5}. 
For larger $\theta$, the spectrum at $\nu=2/5$ is found to be gapless. We conjecture that there are no
incompressible states with $\nu\leq 1/2$ for the interval $\pi/2\leq \theta \leq \pi$.

%%%%%%%%%%%%%%%%%%%%%%%%%%%%%%%%%%%%%%%%%%%%%%%%%%%%%%%%%%%%%%%%%%%%%%%%%%%%%%%%%%%%%%%%%%%%%%

\section{Conclusions}

We have studied the possible quantum Hall states that may be realized with ultracold fermionic
atoms of spin-1/2. This requires first to be able to put all atoms in the LLL for example 
by rotation of the system at low enough temperatures.
Manipulation of the relative strengths of the scattering lengths
allow in principle to probe regimes that are inaccessible to electronic systems. 
When there are attractive interactions in the $p$-wave channel we find that there is phase
separation. The atoms will then form a maximally compact droplet in the presence
of the remaining trapping potential.
The most interesting case is reached when the $p$-wave scattering is repulsive and 
larger than the $s$-wave scattering.
There is then the possibility of paired phases adiabatically connected to a Bose Laughlin fluid
of spin singlet pairs of atoms. 
In principle such phases may be strongly or weakly paired.
We have shown that a paired state is likely to exist at
filling factor $\nu =2/3$ for an extended range of parameters.
When the interaction in the $s$-wave channel is either weakly repulsive or attractive
there is a presumable gapped phase at shift $-3$. It has all the spectroscopic
signatures we expect from a strongly paired state~: there is a magnetoroton branch extending up to $N/2$ values
of the angular momentum. We have also shown that this state has a very good overlap
with a (weakly) paired non-Abelian spin-singlet state which was derived from the (221) Halperin state
in a contruction by Ardonne et al.~\cite{Ardonne02}.

Fine-tuning the scattering properties allows to reach a region
with Coulomb-like interactions where incompressible FQH states are explained readily 
by the composite fermion picture. There is evidence from our studies for an incompressible state
at filling factor $\nu =2/5$ that is explained by wavefunction originally proposed by Halperin.
We have shown also that there are interesting critical states that are already known from
the electronic world that can be reached at some special points in our phase diagram.
These are the so-called Haldane-Rezayi and permanent state. All these findings
give strong motivations to establish and refine experimental techniques that allow to create 
and manipulate ultracold gases in the LLL regime.

\suhe{Acknowledgments} 
One of us (TJ) thank A.~H.~MacDonald for useful discussions. GM acknowledges fruitful dialogs
with N.~R.~Cooper and financial support from EPSRC grant GR/S61263/01 and ESF short-visit grant 
no.~2129 under the INSTANS program.
We thank IDRIS for support and computer time on the large scale computation facility Zahir.
This work is supported in part by ``Institut Francilien des atomes froids'' (IFRAF).

\bibliography{citations}

%%%%%%%%%%%%%%%%%%%%%%%%%%%%%%%%%%%%%%%%%%%%%%%%%%%%%%%%%%%%%%%%%%%%%%%%%%%%%%%%%%%%%%%%%%%%%%%

\end{document}